\documentclass{article}   
\usepackage{natbib}

\usepackage{physics,amsmath,amssymb,mathrsfs}
\usepackage{enumitem}

\title{On efforts to decouple early universe cosmology and quantum gravity phenomenology\footnote{I am grateful for comments and conversations with Nick Huggett and Guilherme Franzmann at various stages, and for correspondence with Anna Ijjas near the end. My reviewers were also helpful, pushing me to clarify several key moments in my exposition and argument. This article was initially drafted and revised while I was a postdoctoral researcher on the Beyond Spacetime project, funded by a `Cosmology Beyond Spacetime' grant from the John Templeton Foundation (no. 61387).}}

\author{Mike D. Schneider\footnote{Department of Philosophy, University of Missouri}}

\date{August 2023}

\begin{document}
\maketitle

\begin{abstract}
The Big Bang singularity in standard model cosmology suggests a program of study in `early universe' quantum gravity phenomenology. Inflation is usually thought to undermine this program's prospects by means of a dynamical diluting argument, but such a view has recently been disputed within inflationary cosmology, in the form of a `trans-Planckian censorship' conjecture. Meanwhile, trans-Planckian censorship has been used outside of inflationary cosmology to motivate alternative early universe scenarios that are tightly linked to ongoing theorizing in quantum gravity. Against the resulting trend toward early universe quantum gravity phenomenology within and without inflation, Ijjas and Steindhardt suggest a further alternative: a `generalized cosmic censorship' principle. I contrast the generalized cosmic censorship principle with the logic of its namesake, the cosmic censorship conjectures. I also remark on foundational concerns in the effective field theory approach to cosmology beyond the standard model, which would be based on that principle.
\end{abstract}


\section{\label{sec:level1}Introduction}

The initial singularity in standard model cosmology suggests a special relationship between early universe cosmology and quantum gravity phenomenology. At sufficiently early moments in cosmic history according to the standard model (`at the Big Bang'), the large-scale cosmos exits a UV physical regime in which we expect methods from semiclassical gravity to fail, including the approximation of quantum gravity states by classical spacetime geometries. One therefore expects empirical traces of UV quantum gravity in the ensuing semiclassical cosmological record. Hence, early universe cosmology serves as a phenomenological window into UV quantum gravity. Call `early universe’ quantum gravity phenomenology the applied theoretical research program that ties together theorizing about UV quantum gravity with empirical work in cosmology, specifically concerning the initial formation or seeds of large-scale cosmic structure amidst mean cosmic expansion. 

The general form of reasoning employed in early universe quantum gravity phenomenology is familiar. It is a logic that is otherwise commonly employed to infer high-energy conditions in the matter sector within early universe epochs, on the basis of evidence in large-scale structure timestamped at later stages of the mean expansion \citep{smeenk2005false,schneider2021trans}. An example of research in early universe quantum gravity phenomenology concerns the Weyl curvature hypothesis, as initially proposed by Penrose \citep{penrose1979singularities} (see also the recent discussion in \citep{kiefer2022quantum}). In this example, cosmological conditions in UV quantum gravity are supposed to select a highly unusual classical spacetime geometry as an empirically accurate semiclassical description of the quantum cosmos, immediately following the Big Bang (i.e. precisely where semiclassical descriptions are thought to become newly viable). 

Still, it is possible that our empirical access to conditions immediately following the Big Bang is itself sufficiently weak as to undermine hopes of treating the semiclassical early universe as a window into UV quantum gravity. Inflation would seem to make this likely: any traces of UV quantum gravity in the cosmos following the Big Bang are substantially diluted through the proposed inflationary epoch, so that whatever evidence we recover today of the early universe is very likely (merely) evidence of conditions during inflation and onward. Inflationary cosmology thereby fosters a \emph{quietism} about early universe quantum gravity phenomenology. Despite generally accepted arguments that inflation will not suffice to change our expectations about the singular structure of the Big Bang \citep{borde2003inflationary}, research efforts within inflationary cosmology are more often focused on issues with inflation's end \citep{guth2007eternal}.\footnote{Certain models of inflation --- famously Starobinsky inflation \citep{starobinsky1980new} --- may be motivated by UV quantum gravity. And the onset of inflation may itself be studied as a dynamical consequence of UV quantum gravity in the very early universe. But I take `inflationary cosmology' to be the research program that approaches any one such model in quantum cosmology from an effective field theory perspective, thereby ignoring underlying UV quantum gravity degrees of freedom as physical causes of low-energy dynamical behavior.}

Regardless of the quietism fostered \emph{within} inflationary cosmology, critics of inflation have consistently challenged the quietism from \emph{without}. As was initially pushed in \citep{martin2001trans}, empirically viable models of inflation would seem to suffer a trans-Planckian problem, so that claims about an inflationary epoch within the early universe would commit one to claims about UV quantum gravity in the underlying quantum cosmos at the onset of inflation as well. The dynamical diluting argument only quiets hopes of UV quantum gravity phenomenology in traces left over from a period prior to inflation; the trans-Planckian problem suggests that, with the onset of inflation, UV quantum gravity grows loud.

In recent years, framing in terms of a trans-Planckian problem has given way to a `trans-Planckian censorship' conjecture \citep{bedroya2019trans,bedroya2020trans}, in light of recent trends in the swampland literature in string theory \citep{palti2019swampland,agrawal2018cosmological,montero2021cobordism,van2022lectures}. Namely, with the onset of inflation, UV quantum gravity would grow too loud, and hence be physically unreasonable: effective field theories that showcase a trans-Planckian problem in cosmology, like many of the empirically viable models of inflation, are thereby conjectured to violate underlying principles of UV quantum gravity. In this respect, empirically successful models of inflation, \emph{and only those which are explicitly shown to be UV quantum gravity compatible}, amount to positive examples of quantum gravity phenomenology in a singular (and inflating) early universe.\footnote{See, e.g., the reasoning on display in \citep{kamali2020warm,bastero2021towards}. For philosophical discussion on the underlying logic of trans-Planckian censorship, including its relationship to the swampland familiar in string theory, see \citep{schneider2021trans,schneider2022strictly}.}

Proponents of trans-Planckian censorship have primarily taken this state of affairs to motivate work on alternative early universe scenarios. (Though that is not to say, in all cases, that the alternative scenarios are immediately known to fare better; failures of trans-Planckian censorship are usually documented in terms of consequences of the theory of cosmological perturbations given mean inflationary expansion, while it is the latter given that is considered variable across the different scenarios.) One of the two initial papers on trans-Planckian censorship was dedicated to this subject, mentioning specifically as alternatives: string gas cosmology, a `Pre-Big Bang' model that exploits a scale factor duality in stringy cosmology analogous to T-duality, and Ekpyrotic scenarios \citep{bedroya2020trans}. Unsurprisingly, these alternative early universe scenarios directly implicate proposed idiosyncratic features of UV quantum gravity in explanations of downstream cosmic structure formation. So one can understand interest in trans-Planckian censorship as all around renewing the prospects of early universe quantum gravity phenomenology.

In light of this recent trend away from quietism, it is noteworthy that \citet{ijjas2018bouncing,ijjas2019new} advocate yet a further alternative: from an effective field theory perspective, a ``classical (non-singular)'' bounce can altogether classically resolve the standard model Big Bang singularity. Resolving the Big Bang singularity classically amounts to a new quietism about early universe quantum gravity phenomenology. But unlike in the case of inflationary cosmology, this new quietism is not primarily motivated by any one dynamical result obtained \emph{within} the particular early universe approach (e.g. where initial traces of UV quantum gravity would be diluted through an ensuing epoch). Instead, the new quietism occurs by fiat of the particular effective field theory framework employed. One supposes that UV quantum gravity is forever screened off from the dynamics of a (bouncing) \emph{effective cosmos}, adopting the position that the empirical study of such an effective cosmos leaves ample room for theorizing later, in some future underlying theory of quantum cosmology, about empirically adequate (non-singular) bounces. Here, I consider the foundations of such a general program of research that is focused on dynamics of an effective cosmos --- a program I dub `cosmology done as effective field theory'.\footnote{\label{fnconstraininglowenergy}Note that here and throughout, I have in mind effective field theory relative to some laboratory frame transported along a comoving or cosmic-stationary worldline within the standard model, as opposed to relative to, e.g., a choice of conformal frame defined over the entire conformal cosmos (see discussion in \citep{ijjas2018space} about the matter Lagrangian being what ultimately disambiguates a `physical' reference scale). As a matter of formalism, effective field theory in the context of general relativistic reasoning within theoretical cosmology faces difficulties \citep{koberinski2022lambda}. But I understand the issues at play to be equally troubling for proponents of inflationary cosmology, the dominant approach in the study of the early universe. On the other hand, like I have already mentioned within inflationary cosmology, trans-Planckian censorship (and also the swampland program) applies pressure to the free employ of the effective field theory framework in quantum cosmology \emph{specifically as providing low-energy descriptions of a quantum cosmos}. These developments in quantum gravity research therefore apply some pressure on a `cosmology done as effective field theory' approach, as well: underlying facts about UV quantum gravity may undermine the empirical adequacy of at least some models of an effective cosmos, when the latter are regarded as low-energy approximations of the quantum cosmos around us. (Though, for a pessimistic take on the pressure that is ultimately applied, see \citep{silk2022swampland}.)} 

\section{Generalized cosmic censorship}

At face value, \citep{ijjas2018bouncing} presents an entirely phenomenological approach to cosmology beyond the standard model: stipulated high-energy dynamics entail effective violations of familiar energy conditions in the vicinity of the Big Bang, so as to evade the scope of the classical Penrose-Hawking-Geroch singularity theorems in general relativity. And the approach evidently enjoys some merits. Not the least, advocates of the approach sidestep troubling conceptual problems with geometrogenesis, or the `flat' emergence of spacetime (supplemented with appropriate conditions on a matter sector), alongside an account of the hierarchical emergence of all the same \citep{crowther2021below}. Indeed, as emphasized throughout \citep{ijjas2018bouncing}, there is, prima facie, no obvious reason to think that satisfying problem-solving in cosmology beyond the standard model cannot perfectly well proceed by means of a phenomenological approach, especially in the context of a particular class of bouncing models that has already been well studied (see references therein). Meanwhile, in the context of that particular class, the approach turns out to handle well familiar puzzles with entropy that are closely related to the Weyl curvature hypothesis mentioned above in the Introduction \citep{ijjas2022entropy}. In short: a phenomenological approach to cosmology beyond the standard model is empirically viable, with attractive notes especially in the context of certain proposals for a bouncing dynamics consistent with effective violations of familiar energy conditions in general relativity. 

In \citep{ijjas2019new}, the same authors consider one particular class of bouncing models in this approach: what they call `the new cyclic cosmology'. As they note in the conclusion there, the new cyclic cosmology witnesses a possible `generalized cosmic censorship' principle. Importantly, the approach to cosmology that would proceed from a generalized cosmic censorship principle is ultimately much broader than just the new cyclic cosmology. Namely, the approach includes all semiclassical cosmologies in which the universe is ``[shielded] from reaching a stage where quantum physics dominates over classical'' [p. 671].\footnote{\label{fnpastincomplete}A modification of the BVG result \citep{borde2003inflationary} is used in \citep{kinney2022cyclic} to argue that the new cyclic cosmology must be past-incomplete with respect to at least some non-comoving geodesic. Often, geodesic incompleteness is treated as something pathological in the classical description, and therefore a signal that quantum gravitational physics comes to dominate over classical somewhere in the vicinity. But signals can be noisy: more work is needed to show that, in the specific case of the past-incomplete geodesics within these cyclic models, it is likely due to some distinctive feature of the underlying quantum gravitational physics (i.e. not to be explained at the level of the effective high-energy dynamics) that the geodesic is rendered incomplete. In \S3 below, I will return to this general topic of the interpretation of incomplete geodesics in cosmology given the generalized cosmic censorship principle, i.e. on a `cosmology done as effective field theory' approach.} (The quoted remark serves, in context, to define their generalized cosmic censorship principle.)

According to generalized cosmic censorship then, UV effects of quantum gravity are everywhere and -when locally screened off from the low-energy scales that characterize semiclassical cosmological modeling from a cosmic-stationary laboratory frame, or cosmology done as effective field theory. Notably, unlike in inflation, interpreted as such an effective field theory dynamics for a specific cosmic epoch \emph{within a singular quantum cosmos}, quantum cosmologies satisfying generalized cosmic censorship allow for an effective field theory interpretation of the entire history of the cosmos. So, per generalized cosmic censorship, we may just as well proceed in cosmology by means of an empirical study of the behavior of an effective cosmos. 

Plausibly, semiclassical cosmologies that would satisfy a generalized cosmic censorship principle include all bouncing resolutions to the Big Bang singularity for which the diagrammatic tool presented in \citep{ijjas2018bouncing} is descriptively apt\footnote{Introducing that diagrammatic tool for use in showcasing the merits of a ``classical (non-singular) bounce'' is the raison d'\^{e}tre for \citep{ijjas2018bouncing} --- interested readers are encouraged to look at the details of the diagrams on their own.} --- thus marrying the phenomenological approach endorsed in that earlier article with a specific way of thinking effective field theoretically about our (ultimately) quantum cosmos. But notably, it is conceivable that there exist, meanwhile, a class of non-bouncing cosmologies that would likewise satisfy a claim of ``classically'' resolving the Big Bang singularity (i.e. consistent with the original author's intended nomenclature). These too would, one presumes, be included under the banner of generalized cosmic censorship.

Here, ``classical'' is a local notion: indicating apt approximations of quantum gravity states by classical spacetime geometry, everywhere and -when in the cosmos --- i.e. about all (quantum) events. The characterization of generalized cosmic censorship in terms of ``shielding'', quoted above, is similarly local in (quantum) spacetime. This is no accident: a Big Bang singularity resolution being classical everywhere and -when in the cosmos is exactly what would shield the universe (otherwise well described by the standard model) from reaching, anywhere and -when, a stage where quantum physics dominates over classical. Likewise, for the universe (otherwise well described by the standard model) to be shielded from reaching a stage where quantum physics dominates over classical, it must be that the Big Bang singularity is resolved (quasi-) locally: the physics responsible for the shielding must kick in precisely in the vicinity of where the large-scale cosmos would otherwise be understood as exiting a UV physical regime --- `at the Big Bang'.

But as the `(quasi)' foreshadows, local notions of Big Bang singularity resolution are tricky to countenance in the context of ordinary thinking about the physical significance of classical global spacetime structure, which precisely includes the topic of spacetime singularities like the Big Bang \citep{earman1995bangs,earman1996tolerance}. In thinking about classical singularity resolution, and therefore also to satisfyingly link classical singularity resolution to a generalized cosmic censorship principle, it is therefore prudent to understand each rather in terms of additional assumptions about the physical fates of information couriers that are confined to propagate about explicit wordlines within possibly singular spacetimes. The topic of information couriers (a term of art introduced here to navigate conversations about possible causal propagation ``through'' classical singularities) is taken up in \S\ref{secinformation} below. Apropos of that discussion, it is plausible that classical resolutions to the Big Bang are semiclassical cosmologies satisfying generalized cosmic censorship, where (moreover) information couriers have optimistic fates though they might otherwise be identified as propagating on past-singular neighborhoods within the standard model.

Whatever is the empirical promise of particular models of an effective cosmos --- including, per the previous paragraph, that they might constitutively guarantee a classical resolution to the standard model, Big Bang singularity --- it is clear that the general approach of empirically studying (only) an effective cosmos rests on the suitability of the generalized cosmic censorship principle. Principles are, of course, familiar in cosmological inquiry. They are applied to our cosmological data in order that we may epistemically secure any number of scientific beliefs, e.g. that all the familiar conclusions of the standard model --- including impressive inferences about the dark sector --- are descriptively apt \citep{smeenk2020some}. So new proposed principles are hardly disqualifying in cosmological inquiry just by their nature. But the question is self-evident: why, moving forward, might we endorse this particular new one? 

One thought is that the generalized cosmic censorship principle is a natural extrapolation from its namesake, the cosmic censorship conjectures. The cosmic censorship conjectures are themselves broadly popular, so modest inferential leaps from their (conjectural) conclusions are perhaps not unreasonable. But the generalized cosmic censorship principle differs from cosmic censorship precisely in virtue of the former's consequence of programmatically endorsing quietism about early universe quantum gravity phenomenology. By contrast, the common thread between the different strengths conjectured of cosmic censorship is usually understood in terms of the paramount importance of predictability in our classical physics, as the latter are to provide approximate descriptions of an ultimately quantum world. The cosmic censorship conjectures constrain the physically viable subset of solutions to the classical dynamics of general relativity, on the basis of underlying considerations about unitary quantum gravity spoiling the formation of classically naked singularities.\footnote{This view is closely connected to the black hole evaporation paradox \citep{hawking1976breakdown,wald2001thermodynamics}. As is pointed out in the ``Open Issues'' section of \citep{wald2001thermodynamics}, if quantum gravity is not unitary, classical gravitational singularities (that witness violations of cosmic censorship) could represent absolute information loss scenarios --- surely an important clue in ongoing quantum gravity research.}

In this regard, the cosmic censorship conjectures are much more similar to the trans-Planckian censorship conjecture mentioned in the Introduction, which was initially proposed in \citep{bedroya2019trans} in connection with work on the quantum gravity swampland in string theory. In the case of trans-Planckian censorship, one conjectures that UV features of quantum gravity in our cosmos place constraints on physically viable effective field theories within quantum cosmology. (The architecture of the swampland, initially developed in string theory, merely makes the constraint claim a bit more precise.) The similar characters of trans-Planckian censorship and cosmic censorship --- that each moves from an anticipated feature of UV quantum gravity to contouring a low-energy landscape of effective dynamics --- indicate that traditional appeals to cosmic censorship in the literature tie descriptions of (black hole) singularities in low-energy regimes directly to underlying facts about UV quantum gravity.\footnote{In fact, motivation for trans-Planckian censorship has been tied directly to familiar thinking about cosmic censorship \citep{brandenberger2021limitations}.} By contrast, the generalized cosmic censorship principle supposes that, at least in the cosmological sector, classical physics ever dominates over its quantum corrections. UV features of quantum gravity in our cosmos thereby \emph{fail to constrain} effective field theory techniques deployed freely in cosmological theorizing. One presumes that nothing in UV quantum gravity spoils the adequacy of those techniques.

Note that it may turn out, in the context of quantum cosmology, that the swampland program (or a trans-Planckian censorship conjecture, or even the cosmic censorship conjectures) constrains \emph{UV quantum gravity compatible models} of an effective cosmos, which are otherwise empirically adequate with respect to cosmological data (cf. footnote \ref{fnconstraininglowenergy}). But the fact that effective models are ultimately beholden in quantum cosmology to underlying facts about UV quantum gravity is not sufficient to regard a model of an effective cosmos, where cosmologically empirically adequate, as describing some quantum gravity phenomena. This is analogous to the commonsense claim that a structural engineering description of a wooden table likewise fails to be a description of quantum gravity phenomena --- despite the underlying facts of the matter.

The basic point is that, whereas trans-Planckian censorship and cosmic censorship are conjectures within and pertinent to ongoing quantum gravity research, the generalized cosmic censorship principle is theoretically inert. While it is quite possible that the cosmos is effective --- like a wooden table, unrevealing of an underlying theory of quantum gravity --- and it is possible that the low-energy landscape of a future quantum gravity theory includes effective dynamics consistent with a classical (non-singular) bounce, the latter possibility would be of little consequence in ongoing quantum gravity research. The only exception would be if we suppose all possible early universe alternatives allowed by the same underlying quantum gravity theory, but which would explicitly draw on UV features of the theory to explain downstream cosmic phenomena, are ruled out on further empirical grounds. (This seems wildly infeasible.) Cosmic censorship and generalized cosmic censorship are therefore crucially dis-analogous. The upshot is that the inferential leap from the conclusions of the former to generalized cosmic censorship would seem not to be modest. The extrapolation of generalized cosmic censorship on the basis of the cosmic censorship conjectures is not so very natural.\footnote{A formal point also illustrates the conceptual gap between cosmic censorship and generalized cosmic censorship. The strongest cosmic censorship conjecture is that spacetime is necessarily globally hyperbolic, validating (by conjecture) 3+1 interpretations of classical general relativity that we might suppose are to be recovered in a suitable limit of the underlying quantum gravity theory. By contrast, it is unclear what interpreted sub-theory (or, indeed, sub-class of effective field theories) we would expect to stand in that same limit, were we to insist on the viability of generalized cosmic censorship in theorizing about the cosmos.}

Still, it is of some interest what such an approach in early universe cosmology would look like, which claims to be forever in cosmic time decoupled from quantum gravity phenomenology, as a matter of underlying principle. In addition to pure scholarly interest in the subject, it also bears mention that, despite enthusiasm surrounding early universe quantum gravity phenomenology within quantum gravity research, a `cosmology done as effective field theory' program might be broadly popular in physics beyond the quantum gravity context. Namely: it represents the cosmological wing of a point of view that physics might very well be considered effectively `all the way down' and across all applications --- or, at least, a point view that we might just as well proceed as if all of this were so, until an abundance of theoretical and empirical evidence jointly compels us to do otherwise.

In the remainder of this article, I consider the `cosmology done as effective field theory' program, which I take to be based on the original discussion found in \citep{ijjas2018bouncing} that precedes the articulation of a generalized cosmic censorship principle. My investigation proceeds from an insight in that earlier article, that an empirically adequate study of an effective cosmos requires a ``classical'' resolution to the Big Bang singularity, i.e. in contrast with alternative semiclassical resolutions.\footnote{I set aside, for the time being, a question as to whether finding leading-order terms dominating within an effective field theory approach to spacetime theories is indeed tantamount to preserving classical spacetime --- thanks to Guilherme Franzmann for flagging this. It may be, of course, that talk of quantum corrections to classical dynamics is simply inappropriate in a quantum cosmology setting.} These alternative resolutions are cosmologies that exhibit (what we might call) \emph{singular} starts or bounces, such that the resolution is itself not classical, and beyond the scope of effective field theory. Examples of singular starts are group field theory condensate cosmology and string gas cosmology (at least, in the absence of finding an effective action that implements it), where there is something like geometrogenesis in the complete quantum cosmological description that yields a  `singular' semiclassical cosmos, i.e. singular at the level of semiclassical description \citep{oriti2021complex}. By contrast, in the case of singular bounces, the semiclassical cosmos would seem, as a matter of phenomenology, to undergo a bounce in the early universe like in the classical cases. But the bounce itself in that description occupies a UV quantum gravity regime. Examples include the initial Ekpyrotic scenarios and Pre-Big-Bang model in string theory (cf. \citep{brandenberger2017bouncing} for discussion of these examples, including references), and perhaps as well the matter bounces studied in the context of loop quantum cosmology \citep{ashtekar2009singularity} or group field theory \citep{oriti2017bouncing}.

In the context of semiclassical cosmologies that possibly include singular starts and bounces as a contrast, it would seem that there are two closely related assumptions about the quasi-local\footnote{`Local' in the sense of neighborhoods; `quasi' because the neighborhoods in question must be, by setup, carefully chosen as opposed to arbitrary.} fundamental physics in the vicinity of a bounce for the diagrammatic tool presented as the main focus in \citep{ijjas2018bouncing} to be apt. These are discussed in the next section, and clarify the constitutive assumptions behind the cosmology done as effective field theory program. In preview and sum: the generalized cosmic censorship principle, which would secure the epistemic legitimacy of the overall program, would seem to be a joint affirmation of these two assumptions about the quasi-local physics throughout an effective cosmos.

\section{Information couriers in spacetime}\label{secinformation}

In \citep{ijjas2018bouncing}, the authors are evidently concerned with bouncing resolutions to a certain class of \emph{nearly Bianchi} cosmologies: globally hyperbolic and singular Big Bang spacetimes that satisfy the dynamics of classical general relativity for various familiar matter sources, which admit Cauchy foliations that approximate homogeneous spatial sections near the initial singularity. As discussed in \citep{bars2014sailing} (by partially overlapping authors), provided that one restricts attention to these settings, one can employ certain conformal techniques to ``lift'' past-incomplete maximal causal geodesics therein, so as to then consider geodesic extensions to those lifted curves. Notably, this procedure can be done to a sufficiently dense set of such geodesics as to cover the underlying spacetime, as one approaches the initial Big Bang singularity. The upshot is that one can think of the extensions to all those lifted curves, taken altogether, as pushing the underlying spacetime that they quasi-locally cover ``beyond'' the Big Bang singularity, and into previous epochs. All it costs is the lift.  

In \cite{carrasco2014journeys}, it is noted that there are still divergent invariants --- hence cosmological singularities of a kind --- in the conformal theory appealed to when performing this lift. But the fact of these divergences does not appear sufficient to undermine the employ of the lift, by means of that conformal theory, to identify geodesic extensions of sufficient numbers of curves into prior epochs. This is a point made in direct reply to \cite{carrasco2014journeys} by \citep[p.3]{bars2014sailing}: ``despite the curvature singularities, physical information can and does journey [...] through the cosmological singularities''. Of course, one may simply find physical divergences troubling, wherever they appear. In this section, I rather develop a different point: a concern about the alternative emphasis on the journeying of physical information \emph{through} such divergences appearing in classical theories thought to be recovered in effective limits within the underlying physics.

As I have just summarized it, the lifting procedure is clearly a formal trick --- not so unlike older formal approaches that associate the Big Bang singularity with definite boundary points (which one might then regard as two-sided, and embedded in something larger). But in the class of cosmologies considered in \citep{ijjas2018bouncing}, which feature a period of slow contraction followed by a bounce, and then by a period of expansion, something suitably similar to this procedure can provide a means of identifying the expanding period with the standard model (that is, despite the latter's singular structure). Hence, one can understand these cosmologies to resolve the Big Bang singularity in the standard model by (quasi-locally) lifting a sufficienty dense set of standard model geodesics near the singularity therein into an alternative setting, which features early contraction followed by a ``classical (non-singular)'' bounce. The diagrammatic tool whose presentation is the focus in \citep{ijjas2018bouncing} thus enters the discussion, as a helpful means of bookkeeping in the ensuing general program of research in cosmology beyond the standard model.

Why discuss techniques to extend already maximal geodesics in singular, Big Bang spacetimes? It may be helpful to recall some basics. The appeals to ``journeying of physical information'' found in \citep{bars2014sailing} are obscure, but the broad point is easy enough to clarify. Typically, causal geodesic curves in a spacetime are identified with the inertial trajectories of test particles through spacetime (where, then, test particles trace out other predictable trajectories, if forced in particular, expected ways --- e.g. by a Lorentz force law, in the case of a charged particle). So, in particular, geodesic curves tell us something about the local communication channels between physical systems that are not spatiotemporally coincident. That is: the geodesic structure of a spacetime describes the viable ideal communication channels linking events therein, as subject to various physical constraints, e.g. whether the courier of the information is massive or massless, charged in the presence of an ambient electromagnetic field, and so on. (Though it may be that, in some cases, the relevant particle limits are poorly motivated as descriptions of the wavefront dynamics of fields in the spacetime, which are to be designated as information couriers \citep{linnemann2021comment}.) From this perspective, maximal geodesics describe the total lifetimes of such idealized couriers, which may pick up information anywhere along their paths, to be dropped off anywhere else.

Consequent to this perspective, in singular cosmologies, which feature incomplete maximal causal geodesics, one is led to conclude that information is lost ``into'' the singularity --- as carried there in finite parameter time by massive or massless couriers that traverse the (respectively timelike or null) incomplete maximal geodesics. In the specific case of the Big Bang singularity in standard model cosmology, every conceivable courier is swallowed up at a finite time toward the past. As has become common in the literature on semiclassical gravity, e.g. in the vicinity of black holes, it would be satisfying to have an answer about where goes the information carried by those couriers, as one reaches the Big Bang (moving toward the past). The foundational point to be emphasized in all of this is that this loss of information into a singularity is ultimately to be understood \emph{quasi-}locally: as a matter of the local wavefront behaviors of courier fields coupled to the spacetime metric within singular neighborhoods --- and this is a wholly classical subject matter \citep{geroch2018motion}. If the couriers' paths could somehow be further extended --- that is, wavefront behaviors of fields in fact extend beyond the confines of the singular spacetime --- then one might hope to trace back the information that comes to be relevant to structure formation in the early universe, all the way to cosmic epochs preceding the Big Bang. 

A question: in the case of bouncing cosmologies, do the techniques to extend incomplete maximal standard model geodesics tell us about viable communication channels between local physical systems pre- and post- Big Bang? Or, in other words, do (a sufficiently dense set of) standard model couriers survive the bounce? In light of the foundational point just raised, I understand this to be a non-trivial technical question, and likely model dependent. But in the context of cosmology done as effective field theory, one assumes outright a positive answer: for ``classical (non-singular) bounces'' as considered by \citep{ijjas2018bouncing}, the accuracy of the leading-order description of the quantum gravity state by classical spacetime geometry ensures, by fiat, at least some couriers' safe passage --- enough to cover the spacetime. Or, at least, this would seem a necessary check that a bouncing model proposed ultimately belongs within the class. It is an assumption within the approach; otherwise, for instance, the diagrammatic tool introduced by \citep{ijjas2018bouncing} would be woefully inaccurate. Numerical relativity demonstrations are helpful here in assessing particular models \citep{cook2020supersmoothing}, to defend the expectation that correlations between the physics pre- and post- bounce are not washed out because of the fates of the information couriers in between. 

Meanwhile, there is a more profound way to ruin the fates of the couriers in the semiclassical model, which is motivated by considering singular bounces. Consider the scenario where, in the vicinity of the bounce, coarse-graining spacetime (i.e. coarse-graining over the semiclassical approximation of the quantum cosmos) fails to ``commute'' with the spacetime description (i.e. semiclassical approximation) of the symmetry-reduced (i.e. coarse-grained) quantum cosmos. This is a situation where geodesic structure is likely unfaithful to the behaviors of information couriers understood fundamentally in terms of quasi-locally propagating fields.

A further assumption undergirding the bouncing cosmologies diagrammed in  \citep{ijjas2018bouncing} is therefore that there is a global well-behavedness between the coarse-grained metric structure of the bouncing cosmology depicted in the diagram and the micro-causal structure of the physics that is relevant quasi-locally throughout, and in particular through the bounce. Without such an assumption, those communication channels that are taken to classically permit a courier's safe passage through the bounce in the effective description need not hug the paths that the information couriers are ultimately compelled to follow. 

On this matter, it is important to note that, conceivably, certain classical spacetime descriptions taken as correct to leading order locally with respect to the as-yet unknown theory of quantum gravity (i.e. as a matter of dynamics) may nonetheless violate this global assumption. And so, it is not the mere demonstration that cosmic history is always locally at energy scales well below the Planck-scale that renders the bouncing cosmologies considered by \citep{ijjas2018bouncing} satisfying resolutions to the standard model Big Bang. As a cheap example, consider the following spacetime construction, where coarse-graining procedures obscure the presence of Cauchy horizons in spacetimes with otherwise non-pathological global structure.

The `top half of Misner' spacetime is a particular flat, two-dimensional spacetime diffeomorphic to $S\times\mathbb{R}$; a `top half' because the line element, when written globally as $ds^2=-t^{-1}dt^2+t(d\phi)^2$, admits $\phi$ over the usual complete range $[0,2\pi)$, but $t$ only over $(0,\infty)$. Maximal causal geodesics are incomplete in the top half of Misner spacetime, so that $t=0$ will be our stand-in for a cosmological singularity. Unlike a cosmological singularity, however, the top half of Misner spacetime is smoothly extendable. With a coordinate change $\phi'=\phi+\log t$, the spacetime may be seen as properly isometrically embedded in another, `Misner' spacetime, whose line element may be written globally as $ds^2=-2d\phi'dt+t(d\phi')^2$ for $t$ now ranging over all Real values. At $t=0$ in Misner spacetime, the `top half of Misner spacetime' merely evolves a Cauchy horizon \citep{hawking1973large}. The intuitive picture is of a lightcone in the top half toppling over as one approaches $t=0$ from above, with closed timelike curves immediately beyond. The underlying manifold is nonetheless free of pathology: although there remain geodesics from the top half that are past incomplete in the extension to Misner spacetime, there is also a congruence of causal geodesics in the top half that extend as a congruence of curves in the complete Misner spacetime. Like in the discussion above, extensions of sufficient geodesics identified in the original spacetime would seem to `journey' through the relevant stand-in for a cosmological singularity. 

But where to might those extensions journey? Due to the presence of the Cauchy horizon at $t=0$ in Misner spacetime, it is easy to generate many other (i.e. non-isometric) spacetime extensions to the original top half. For instance, consider two copies of the restriction of Misner spacetime to $t>\epsilon$, for some $\epsilon<0$. (Clearly, each copy properly embeds the top half of Misner spacetime.) Inverting the second copy (`turning it upside down'), one can imagine gluing it to the first at a $t=\epsilon$ seam, using smooth operations within a neighborhood of that seam to ensure that the resulting construction is a spacetime. As long as the chosen neighborhood of the seam is disjoint with the regions of the two copies that each properly embeds the top half of Misner spacetime, the resulting spacetime will suffice for our purposes. Intuitively: the lightcone in the top half, having already toppled over as one approaches $t=0$ from above, at some point thereafter spontaneously begins to right itself in the coordinate range of $2\epsilon<t<0$. The metric as $t\rightarrow -\infty$ in the resulting construction is identical to the metric as $t\rightarrow\infty$ in the top half, and the spacetime contains a Cauchy horizon in two components, with a causally bizarre throat formed between those components. (Meanwhile, just like in Misner spacetime, although many causal curves in the top half remain incomplete in the extension, one can find a congruence of curves that extends a congruence of causal curves from the top half.)

This construction is totally general for t-coordinate values $\epsilon<0$ in Misner spacetime. So suppose $\epsilon$ is very small, i.e. $\epsilon\approx 0$ relative to a choice of scale imposed on the resulting construction. Finding a coarse-grained expression of the spacetime metric at that scale, except when done delicately, will miss that the lightcone in the top half region has ever fully toppled over, before proceeding to right itself as $t \rightarrow -\infty$. In this way, one might overlook, in virtue of the coarse-graining scale, the presence of the causally bizarre throat: that there is a region, in the vicinity of the stand-in for the cosmological singularity in the top half, within which the micro-causal structure of the constructed spacetime is utterly perverse. In such a case, geodesic structure specified as a coarse-grained description of the spacetime would seem patently ill suited for considerations about viable communication channels between physical systems at $t \rightarrow \infty$ and $t \rightarrow -\infty$. The couriers that hug those geodesics away from the two-component Cauchy horizon behave wildly differently in the throat formed between the horizon's two components. (Indeed, that the throat is causally bizarre could suggest that it becomes inappropriate to talk about information couriers persisting at all through the region.) The upshot is that extending sufficient numbers of incomplete maximal geodesic curves in a coarse-graining of the top half of Misner spacetime into a coarse-graining of this spacetime construction winds up rampantly unfaithful to the physical propagation of information within the underlying construction, exactly in the vicinity of the singularity in the top half.

\section{One last remark}

Cosmology done as effective field theory is a viable approach in theoretical cosmology beyond the standard model. And some cosmological models consistent with the approach, like the new cyclic cosmology studied by \citet{ijjas2019new}, provide routes to solving at least some familiar problems with the Big Bang and singular starts or bounces \citep{ijjas2022entropy}. But as flagged already in footnote \ref{fnconstraininglowenergy}, cosmology done as effective field theory does not quite succeed in fully decoupling cosmology and quantum gravity research. For instance, trans-Planckian censorship has been discussed in relation to an aspect of the new cyclic cosmology other than the bounce: the end of de Sitter-like expansion within each cycle, as relates especially to solving the coincidence problem familiar in the standard model \citep{scherrer2019coincidence,andrei2022rapidly}. So, prior confidences in particular models within the approach relative to others can be influenced by conjectures like trans-Planckian censorship within quantum gravity research. Still, like in the case of a wooden table, it is difficult to move from this `weak’ coupling of research topics to anything like quantum gravity phenomenology.

Generally, one might wonder as to the virtues of committing, as a matter of principle, to an approach in cosmology beyond the standard model that would spoil prospects in early universe quantum gravity phenomenology. While there are certain procedural advantages in quarantining regimes of ignorance from our scientific theorizing elsewhere, doing so here does cut off an otherwise exciting intersection in research, which promises increased empirical traction on a physical regime in which it is famously difficult to get any substantive empirical traction. In the philosophical literature, `methodological conservatism' refers to the position that we hold onto established scientific beliefs until we are compelled by evidence to drop them \citep{sklar1975methodological}. But not all beliefs relevant to scientific theorizing are themselves scientific beliefs. In particular, the belief that it suffices to pursue theoretical cosmology by means of studying dynamics of an effective cosmos is not itself a scientific belief. True, such a belief has, perhaps, not previously led us astray. And indeed, despite clues from the singular structure of the standard model, the cosmos may in fact be as unrevealing of underlying quantum gravity as is an ordinary wooden table. \emph{Some familiar cosmological problems might even be conceptually easier to solve, given such a belief.} Yet still, it strikes me that there is little reason to newly embrace generalized cosmic censorship as a matter of principle, moving forward.



\bibliographystyle{chicago}
\bibliography{effectivecosmos}

\begin{thebibliography}{}

\bibitem[\protect\citeauthoryear{Agrawal, Obied, Steinhardt, and Vafa}{Agrawal
  et~al.}{2018}]{agrawal2018cosmological}
Agrawal, P., G.~Obied, P.~J. Steinhardt, and C.~Vafa (2018).
\newblock On the cosmological implications of the string swampland.
\newblock {\em Physics Letters B\/}~{\em 784}, 271--276.

\bibitem[\protect\citeauthoryear{Andrei, Ijjas, and Steinhardt}{Andrei
  et~al.}{2022}]{andrei2022rapidly}
Andrei, C., A.~Ijjas, and P.~J. Steinhardt (2022).
\newblock Rapidly descending dark energy and the end of cosmic expansion.
\newblock {\em Proceedings of the National Academy of Sciences\/}~{\em
  119\/}(15), e2200539119.

\bibitem[\protect\citeauthoryear{Ashtekar}{Ashtekar}{2009}]{ashtekar2009singularity}
Ashtekar, A. (2009).
\newblock Singularity resolution in loop quantum cosmology: a brief overview.
\newblock In {\em Journal of Physics: Conference Series}, Volume 189, pp.\
  012003. IOP Publishing.

\bibitem[\protect\citeauthoryear{Bars, Steinhardt, and Turok}{Bars
  et~al.}{2014}]{bars2014sailing}
Bars, I., P.~Steinhardt, and N.~Turok (2014).
\newblock Sailing through the big crunch-big bang transition.
\newblock {\em Physical Review D\/}~{\em 89\/}(6), 061302.

\bibitem[\protect\citeauthoryear{Bastero-Gil, Berera, Ramos, and
  Rosa}{Bastero-Gil et~al.}{2021}]{bastero2021towards}
Bastero-Gil, M., A.~Berera, R.~O. Ramos, and J.~G. Rosa (2021).
\newblock Towards a reliable effective field theory of inflation.
\newblock {\em Physics Letters B\/}~{\em 813}, 136055.

\bibitem[\protect\citeauthoryear{Bedroya, Brandenberger, Loverde, and
  Vafa}{Bedroya et~al.}{2020}]{bedroya2020trans}
Bedroya, A., R.~Brandenberger, M.~Loverde, and C.~Vafa (2020).
\newblock Trans-planckian censorship and inflationary cosmology.
\newblock {\em Physical Review D\/}~{\em 101\/}(10), 103502.

\bibitem[\protect\citeauthoryear{Bedroya and Vafa}{Bedroya and
  Vafa}{2020}]{bedroya2019trans}
Bedroya, A. and C.~Vafa (2020).
\newblock Trans-planckian censorship and the swampland.
\newblock {\em Journal of High Energy Physics\/}~{\em 2020\/}(9), 1--34.

\bibitem[\protect\citeauthoryear{Borde, Guth, and Vilenkin}{Borde
  et~al.}{2003}]{borde2003inflationary}
Borde, A., A.~H. Guth, and A.~Vilenkin (2003).
\newblock Inflationary spacetimes are incomplete in past directions.
\newblock {\em Physical review letters\/}~{\em 90\/}(15), 151301.

\bibitem[\protect\citeauthoryear{Brandenberger}{Brandenberger}{2021}]{brandenberger2021limitations}
Brandenberger, R. (2021).
\newblock Limitations of an effective field theory treatment of early universe
  cosmology.
\newblock {\em arXiv preprint arXiv:2108.12743\/}.

\bibitem[\protect\citeauthoryear{Brandenberger and Peter}{Brandenberger and
  Peter}{2017}]{brandenberger2017bouncing}
Brandenberger, R. and P.~Peter (2017).
\newblock Bouncing cosmologies: progress and problems.
\newblock {\em Foundations of Physics\/}~{\em 47\/}(6), 797--850.

\bibitem[\protect\citeauthoryear{Carrasco, Chemissany, and Kallosh}{Carrasco
  et~al.}{2014}]{carrasco2014journeys}
Carrasco, J. J.~M., W.~Chemissany, and R.~Kallosh (2014).
\newblock Journeys through antigravity?
\newblock {\em Journal of High Energy Physics\/}~{\em 2014\/}(1), 1--13.

\bibitem[\protect\citeauthoryear{Cook, Glushchenko, Ijjas, Pretorius, and
  Steinhardt}{Cook et~al.}{2020}]{cook2020supersmoothing}
Cook, W.~G., I.~A. Glushchenko, A.~Ijjas, F.~Pretorius, and P.~J. Steinhardt
  (2020).
\newblock Supersmoothing through slow contraction.
\newblock {\em Physics Letters B\/}~{\em 808}, 135690.

\bibitem[\protect\citeauthoryear{Crowther}{Crowther}{2021}]{crowther2021below}
Crowther, K. (2021).
\newblock As below, so before:‘synchronic’and ‘diachronic’conceptions
  of spacetime emergence.
\newblock {\em Synthese\/}~{\em 198\/}(8), 7279--7307.

\bibitem[\protect\citeauthoryear{Earman}{Earman}{1995}]{earman1995bangs}
Earman, J. (1995).
\newblock {\em Bangs, crunches, whimpers, and shrieks: Singularities and
  acausalities in relativistic spacetimes}.
\newblock Oxford University Press.

\bibitem[\protect\citeauthoryear{Earman}{Earman}{1996}]{earman1996tolerance}
Earman, J. (1996).
\newblock Tolerance for spacetime singularities.
\newblock {\em Foundations of Physics\/}~{\em 26\/}(5), 623--640.

\bibitem[\protect\citeauthoryear{Geroch and Weatherall}{Geroch and
  Weatherall}{2018}]{geroch2018motion}
Geroch, R. and J.~O. Weatherall (2018).
\newblock The motion of small bodies in space-time.
\newblock {\em Communications in Mathematical Physics\/}~{\em 364\/}(2),
  607--634.

\bibitem[\protect\citeauthoryear{Guth}{Guth}{2007}]{guth2007eternal}
Guth, A.~H. (2007).
\newblock Eternal inflation and its implications.
\newblock {\em Journal of Physics A: Mathematical and Theoretical\/}~{\em
  40\/}(25), 6811.

\bibitem[\protect\citeauthoryear{Hawking}{Hawking}{1976}]{hawking1976breakdown}
Hawking, S.~W. (1976).
\newblock Breakdown of predictability in gravitational collapse.
\newblock {\em Physical Review D\/}~{\em 14\/}(10), 2460.

\bibitem[\protect\citeauthoryear{Hawking and Ellis}{Hawking and
  Ellis}{1973}]{hawking1973large}
Hawking, S.~W. and G.~F.~R. Ellis (1973).
\newblock {\em The large scale structure of space-time}, Volume~1.
\newblock Cambridge university press.

\bibitem[\protect\citeauthoryear{Ijjas}{Ijjas}{2018}]{ijjas2018space}
Ijjas, A. (2018).
\newblock Space-time slicing in horndeski theories and its implications for
  non-singular bouncing solutions.
\newblock {\em Journal of Cosmology and Astroparticle Physics\/}~{\em
  2018\/}(02), 007.

\bibitem[\protect\citeauthoryear{Ijjas and Steinhardt}{Ijjas and
  Steinhardt}{2018}]{ijjas2018bouncing}
Ijjas, A. and P.~J. Steinhardt (2018).
\newblock Bouncing cosmology made simple.
\newblock {\em Classical and Quantum Gravity\/}~{\em 35\/}(13), 135004.

\bibitem[\protect\citeauthoryear{Ijjas and Steinhardt}{Ijjas and
  Steinhardt}{2019}]{ijjas2019new}
Ijjas, A. and P.~J. Steinhardt (2019).
\newblock A new kind of cyclic universe.
\newblock {\em Physics Letters B\/}~{\em 795}, 666--672.

\bibitem[\protect\citeauthoryear{Ijjas and Steinhardt}{Ijjas and
  Steinhardt}{2022}]{ijjas2022entropy}
Ijjas, A. and P.~J. Steinhardt (2022).
\newblock Entropy, black holes, and the new cyclic universe.
\newblock {\em Physics Letters B\/}~{\em 824}, 136823.

\bibitem[\protect\citeauthoryear{Kamali, Motaharfar, and Ramos}{Kamali
  et~al.}{2020}]{kamali2020warm}
Kamali, V., M.~Motaharfar, and R.~O. Ramos (2020).
\newblock Warm brane inflation with an exponential potential: a consistent
  realization away from the swampland.
\newblock {\em Physical Review D\/}~{\em 101\/}(2), 023535.

\bibitem[\protect\citeauthoryear{Kiefer}{Kiefer}{2022}]{kiefer2022quantum}
Kiefer, C. (2022).
\newblock On a quantum weyl curvature hypothesis.
\newblock {\em AVS Quantum Science\/}~{\em 4\/}(1), 015607.

\bibitem[\protect\citeauthoryear{Kinney and Stein}{Kinney and
  Stein}{2022}]{kinney2022cyclic}
Kinney, W.~H. and N.~K. Stein (2022).
\newblock Cyclic cosmology and geodesic completeness.
\newblock {\em Journal of Cosmology and Astroparticle Physics\/}~{\em
  2022\/}(06), 011.

\bibitem[\protect\citeauthoryear{Koberinski and Smeenk}{Koberinski and
  Smeenk}{2022}]{koberinski2022lambda}
Koberinski, A. and C.~Smeenk (2022).
\newblock $\lambda$ and the limits of effective field theory.
\newblock {\em Philosophy of Science\/}, 1--26.

\bibitem[\protect\citeauthoryear{Linnemann and Read}{Linnemann and
  Read}{2021}]{linnemann2021comment}
Linnemann, N. and J.~Read (2021).
\newblock Comment on ‘do electromagnetic waves always propagate along null
  geodesics?’.
\newblock {\em Classical and Quantum Gravity\/}~{\em 38\/}(23), 238001.

\bibitem[\protect\citeauthoryear{Martin and Brandenberger}{Martin and
  Brandenberger}{2001}]{martin2001trans}
Martin, J. and R.~H. Brandenberger (2001).
\newblock Trans-planckian problem of inflationary cosmology.
\newblock {\em Physical Review D\/}~{\em 63\/}(12), 123501.

\bibitem[\protect\citeauthoryear{Montero and Vafa}{Montero and
  Vafa}{2021}]{montero2021cobordism}
Montero, M. and C.~Vafa (2021).
\newblock Cobordism conjecture, anomalies, and the string lamppost principle.
\newblock {\em Journal of High Energy Physics\/}~{\em 2021\/}(1), 1--47.

\bibitem[\protect\citeauthoryear{Oriti}{Oriti}{2021}]{oriti2021complex}
Oriti, D. (2021).
\newblock The complex timeless emergence of time in quantum gravity.
\newblock {\em arXiv preprint arXiv:2110.08641\/}.

\bibitem[\protect\citeauthoryear{Oriti, Sindoni, and Wilson-Ewing}{Oriti
  et~al.}{2017}]{oriti2017bouncing}
Oriti, D., L.~Sindoni, and E.~Wilson-Ewing (2017).
\newblock Bouncing cosmologies from quantum gravity condensates.
\newblock {\em Classical and Quantum Gravity\/}~{\em 34\/}(4), 04LT01.

\bibitem[\protect\citeauthoryear{Palti}{Palti}{2019}]{palti2019swampland}
Palti, E. (2019).
\newblock The swampland: introduction and review.
\newblock {\em Fortschritte der Physik\/}~{\em 67\/}(6), 1900037.

\bibitem[\protect\citeauthoryear{Penrose}{Penrose}{1979}]{penrose1979singularities}
Penrose, R. (1979).
\newblock Singularities and time-asymmetry.
\newblock {\em General Relativity: An Einstein centenary survey\/}, 581--638.

\bibitem[\protect\citeauthoryear{Scherrer}{Scherrer}{2019}]{scherrer2019coincidence}
Scherrer, R.~J. (2019).
\newblock The coincidence problem and the swampland conjectures in the
  ijjas-steinhardt cyclic model of the universe.
\newblock {\em Physics Letters B\/}~{\em 798}, 134981.

\bibitem[\protect\citeauthoryear{Schneider}{Schneider}{2021}]{schneider2021trans}
Schneider, M.~D. (2021).
\newblock Trans-planckian philosophy of cosmology.
\newblock {\em Studies in History and Philosophy of Science Part A\/}~{\em 90},
  184--193.

\bibitem[\protect\citeauthoryear{Schneider}{Schneider}{2022}]{schneider2022strictly}
Schneider, M.~D. (2022).
\newblock A (strictly) contemporary perspective on trans-planckian censorship.
\newblock {\em Foundations of Physics\/}~{\em 52\/}(4), 76.

\bibitem[\protect\citeauthoryear{Silk and Cass{\'e}}{Silk and
  Cass{\'e}}{2022}]{silk2022swampland}
Silk, J. and M.~Cass{\'e} (2022).
\newblock Swampland revisited.
\newblock {\em Foundations of Physics\/}~{\em 52\/}(4), 1--11.

\bibitem[\protect\citeauthoryear{Sklar}{Sklar}{1975}]{sklar1975methodological}
Sklar, L. (1975).
\newblock Methodological conservatism.
\newblock {\em The Philosophical Review\/}~{\em 84\/}(3), 374--400.

\bibitem[\protect\citeauthoryear{Smeenk}{Smeenk}{2005}]{smeenk2005false}
Smeenk, C. (2005).
\newblock False vacuum: Early universe cosmology and the development of
  inflation.
\newblock In {\em The universe of general relativity}, pp.\  223--257.
  Springer.

\bibitem[\protect\citeauthoryear{Smeenk}{Smeenk}{2020}]{smeenk2020some}
Smeenk, C. (2020).
\newblock Some reflections on the structure of cosmological knowledge.
\newblock {\em Studies in History and Philosophy of Science Part B: Studies in
  History and Philosophy of Modern Physics\/}~{\em 71}, 220--231.

\bibitem[\protect\citeauthoryear{Starobinsky}{Starobinsky}{1980}]{starobinsky1980new}
Starobinsky, A.~A. (1980).
\newblock A new type of isotropic cosmological models without singularity.
\newblock {\em Physics Letters B\/}~{\em 91\/}(1), 99--102.

\bibitem[\protect\citeauthoryear{van Beest, Calder{\'o}n-Infante,
  Mirfendereski, and Valenzuela}{van Beest et~al.}{2022}]{van2022lectures}
van Beest, M., J.~Calder{\'o}n-Infante, D.~Mirfendereski, and I.~Valenzuela
  (2022).
\newblock Lectures on the swampland program in string compactifications.
\newblock {\em Physics Reports\/}~{\em 989}, 1--50.

\bibitem[\protect\citeauthoryear{Wald}{Wald}{2001}]{wald2001thermodynamics}
Wald, R.~M. (2001).
\newblock The thermodynamics of black holes.
\newblock {\em Living reviews in relativity\/}~{\em 4\/}(1), 1--44.

\end{thebibliography}

\end{document}